 \documentstyle[prl,aps,multicol,epsfig]{revtex}
\begin{document}
\draft \date{\today} \title
  {Meissner-London currents in superconductors with rectangular
   cross section}

\author{Ernst Helmut Brandt}
\address{Max-Planck-Institut f\"ur Metallforschung,
   D-70506 Stuttgart, Germany}
\author{Grigorii P.~Mikitik}
  \address{B.~Verkin Institute for Low Temperature Physics \&
   Engineering, National Ukrainian Academy of Sciences,
   Kharkov 310164, Ukraine}
\maketitle

\begin{abstract}
  Exact analytic solutions are presented for the magnetic moment
and screening currents in the Meissner state of superconductor
strips with rectangular cross section in a perpendicular magnetic
field and/or with transport current. The extension to finite London
penetration is achieved by an elegant numerical method which works
also for disks.
The surface current in the specimen corners diverges as $l^{-1/3}$
where l is the distance from the corner. This enhancement reduces
the barrier for vortex penetration and should increase the
nonlinear Meissner effect in $d$-wave superconductors.
\end{abstract}
    \begin{multicols}{2}
    \narrowtext

   The main feature of superconductors is that they expel weak
magnetic fields $H$ from their interior. This Meissner effect was
described quantitatively by the London brothers, who showed that
$H$ penetrates exponentially to the London depth $\lambda$ \cite{1},
and by Ginzburg and Landau and by Pippard, who introduced
the superconducting coherence length $\xi$. In extreme type-II
superconductors with $\lambda \gg \xi$, the correction caused by
finite $\xi$ to the Meissner-London state usually may be
disregarded, but now the penetration of magnetic flux in form of
Abrikosov vortices has to be considered. Vortex penetration is
governed by surface barriers which increase with decreasing $\xi$.
Both the microscopic Bean-Livingstom barrier \cite{2,3} and the
macroscopic geometric barrier \cite{4,5,6,7} depend on the surface
screening currents flowing in the Meissner-London state. These
currents crucially depend on the specimen shape and on the
orientation of the applied magnetic field $H$. The screening
current is particularly large near sharp edges, where it causes a
reduction of the field of first penetration of vortices and
possibly increases the nonlinear Meissner effect in $d$-wave
superconductors \cite{8,9}. Knowledge of the Meissner-London state
of thin platelets is also required for a correct evaluation of
certain precision measurements of the London penetration depth
$\lambda$ \cite{9}.

    In spite of its fundamental nature, exact solutions of London
theory exist only for the trivial (and less important for
experiments) longitudinal geometry of infinite slabs and
cylinders in parallel $H$, and for the sphere and infinite
cylinder in arbitrarily oriented $H$ \cite{1}. Even for the ideal
Meissner state, i.e.\ the limit $\lambda \to 0$, the only nontrivial
solution we know of is the ellipsoid \cite{10,11} and strips with
elliptic \cite{12} and oval \cite{13} cross sections. In general,
when $\lambda$ is much smaller than the smallest extension of the
superconductor, the surface screening current is
$J = |H_\| |$ where $H_\|$  is the component of ${\bf H(r)}$ at and
parallel to the surface, and its density is approximately
  \begin{eqnarray}    
    j = (J/\lambda) \exp(-\delta / \lambda ) \,,
  \end{eqnarray}
where $\delta$ is the distance from the surface.

 In the present paper we first give an exact solution for the surface
current and magnetic moment in the ideal Meissner state ($\lambda=0$)
of an infinitely long strip with rectangular cross section in a
perpendicular magnetic field $H$ or with transport current $I$.
We then show how the Meissner-London state of this strip (and of
circular disks) can be computed for arbitrary $\lambda$. In all
these cases the solutions for $H>0$, $I=0$ and $H=0$, $I>0$ may
be superimposed linearly to give the general (less symmetric)
result for simultaneously applied $H$ and $I$.
The rectangular cross section of the strip fills the area
$-a \le x\le a$, $-b \le y \le b$; the currents flow along the
strip $\| z$, and $H$ is applied along $y$, see inset in Fig.~1.
The result for arbitrary orientation of $H$ is obtained by
linear superposition of the solutions with $H \|y$ and $H\| x$.

   The surface currents of an ideally diamagnetic rectangular
strip in a perpendicular field $H \| y$ or with transport current
$ I \| z$ can be calculated by conformal mapping. We give here the
main results starting with the case $H>0$, $I=0$; details will
be published elsewhere. Defining the function
  \begin{eqnarray}    
  f(s,m)=ms \int_0^1 {\sqrt{1-s^2t^2} \over \sqrt{1-ms^2t^2}}\,dt
  \end{eqnarray}
[$f$ is the sum of two incomplete elliptic integrals,
$f(s,m) =E(\theta,k) -(1-k^2)F(\theta,k)$, $s=\sin\theta$, $m=k^2$,
$0 \le s \le 1$, $0 \le m \le 1$] we may write the screening
currents and the magnetic moment in parametric form, with curve
parameters $s$ and $m$. First we find $m(b/a)$ from
  \begin{eqnarray}    
  {b \over a} = {f(1,m) \over f(1, 1-m)}
  \end{eqnarray}
and then the currents as functions of $x(s)$ and $y(s)$:
  \begin{eqnarray}    
  {J(x,b) \over H} = {s \over \sqrt{1 -s^2} }, ~~~
   {x \over a} = {f(s,1-m) \over f(1, 1-m)} \,,
  \end{eqnarray}
  \begin{eqnarray}    
  {J(a,y) \over H} = {\sqrt{1-ms^2} \over \sqrt{m(1 -s^2)}},~~~
   {y \over b} = {f(s,m) \over f(1,m)} \,.
  \end{eqnarray}
The (negative) magnetic moment of the strip per unit length along
$z$ is
  \begin{eqnarray}    
  {-M \over \pi a^2 H} = {1-m \over [f(1, 1-m)]^2}
  \end{eqnarray}
with $m=m(b/a)$ from Eq.\ (3). These exact results are shown in
Figs.~1 and 2.

    Some useful approximations and limiting cases are as follows.
For all aspect ratios $0 < b/a < \infty$ one has
  \begin{eqnarray}    
  m= {1 \over 2} +{1\over 2} \tanh\Big(0.463\ln{b\over a} \Big)
      +\epsilon \,,
  \end{eqnarray}
with error $|\epsilon | \le 0.18 \%$.
The exact limits are $m \to 4b/\pi a$ ($b \ll a$) and
$m \to 1 - 4a/\pi b$ ($b \gg a$). The magnetic moment (6)
for all ratios $b/a$ is well fitted by
  \begin{eqnarray}    
  {-M \over (\pi a^2 +4ab) H} =
   1 +\exp\Big(1.6875 -\sqrt{p} \Big) -\epsilon  \,,
  \end{eqnarray}
with $p = 10 +| \ln(b/a) +0.288 |^{1.968} $ and deviation
$0 \le \epsilon \le 0.004$, see Fig.~2.
Formula (8) is more accurate than the fit given in
Ref.\ \onlinecite{14}. The exact limits are (Fig.~2):
  \begin{eqnarray}    
  {-M \over \pi a^2 H}&=& 1+{2b\over \pi a}\Big( \ln{4\pi a\over b}
        -1 \Big), ~~~~ b \ll a \,,  \\
  {-M \over 4 a b H } &=& 1+{a\over \pi b}\Big( \ln{4\pi b\over a}
        +{1 \over 2} \Big), ~~~~ b \gg a \,.
  \end{eqnarray}

   The surface currents diverge at the corners symmetrically
as $l^{-1/3}$ where $l$ is the distance from the corner,
e.g.\ $l=a-x$ and $l=b-y$. Near the corners one has
  \begin{eqnarray}    
     J_{\rm corner} = H \left[ {(1-m) \over 3 \sqrt{m} f(1,m)}
     \, {b \over l} \right]^{1/3} .
  \end{eqnarray}
At the equator  $J(a,0)= H / \sqrt{m}$ holds and near the poles
$J(x,b)/H \approx (x/a)f(1,1-m)/(1-m)$ which equals $\pi x/4a$
at $b \gg a$ and $x/a$ at $b \ll a$.
For long slabs one has $J(a,y) \approx H$ except near
the corners [see Eq.\ (11)]. For thin strips ($b\ll a$) Eq.~(4)
yields
  \begin{eqnarray}    
  J(x,b) = H\, x\, (a^2 - x^2)^{-1/2}
  \end{eqnarray}
in agreement with the known sheet current $2J(x)$ \cite{15,16}.
On the edge of thin strips one has the fit
  \begin{eqnarray}    
  J(a,y) = H m^{-1/2} \, [1 - (y/b)^2 ]^{-0.31}
  \end{eqnarray}
with relative error $< 1\%$ for $y/b < 0.92$ or
$<2\%$ for $y/b < 0.97$.
More limiting expressions and fits are easily derived from the
exact Eqs.~(2-6).

  For the strip with current $I>0$ and $H=0$ we find the
Meissner surface currents
  \begin{eqnarray}    
   J(x,b) ={I\over 2\pi a} {f(1,1-m) \over
            \sqrt{1-m}\sqrt{1 -s^2} }, ~~
   {x \over a} = {f(s,1-m) \over f(1, 1-m)} \,,
  \end{eqnarray}
  \begin{eqnarray}    
   J(a,y) ={I\over 2\pi b} {f(1,m) \over
            \sqrt{m(1 -s^2) } }, ~~~
   {y \over b} = {f(s,m) \over f(1,m)} \,.
  \end{eqnarray}
These expressions are invariant when interchanging
$a, b$ and  $x, y$, which replaces $m$ by $1-m$. In the corners
the current again diverges as $l^{-1/3}$, cf.\ Eq.~(11):
  \begin{eqnarray}    
     J_{\rm corner} = {I \over 2\pi b} \left[ {f(1,m)^2 \over
      3\sqrt{m-m^2}} \, {b \over l} \right]^{1/3} .
  \end{eqnarray}

  Accounting for a small but finite London depth $\lambda \ll a,b$,
the screening currents penetrate exponentially from the surface,
e.g., $j(x,y) = J(x,b)\lambda^{-1} \exp[(y-b)/\lambda]$ near $y =b$.
The magnetic moment $M =\int\! dx \int\! dy\, x\, j(x,y)$
is thus slightly reduced to approximately
  \begin{eqnarray}    
  M(a,b,\lambda) \approx M(a-\lambda, b-\lambda, 0)
  \end{eqnarray}
with $M(a,b,0)$ from Eqs.~(6, 8-10). Similar formulas
are valid for specimens of any shape if the radius of curvature
of the surface is much larger than $\lambda$. For our
strip, however, the sharp rectangular corners give an additional
contribution $\delta M_{\rm corner}$ to $M(\lambda) - M(0)$.
When $[ \lambda / {\rm min}(a,b) ]^{1/3} \ll 1$, one has
from Eq.~(11)
  \begin{eqnarray}    
  \delta M_{\rm corner} \propto (\lambda^2 /ab)^{1/3} M \,.
  \end{eqnarray}
This nonanalytical term  dominates in
$\partial M/\partial \lambda$ and may explain some experimental
findings in Ref.~\onlinecite{9}.
   In the opposite case, $\lambda \gg a,b$, the vector potential
of the induced current density is negligible,
thus $j \approx -Hx/\lambda^2$  and
  \begin{eqnarray}    
  M(a,b,\lambda) \approx -(4a^3 b/3 \lambda^2)  H \,.
  \end{eqnarray}

   Next we show how the current density $j(x,y)$ and magnetic
moment $M$ of thick strips (and of disks) for finite London depth
$\lambda$ can be obtained in an elegant way, avoiding the
calculation and cutoff \cite{17} of the magnetic field around the
strip. The static London equation reads
  \begin{eqnarray}    
   -\lambda^2 \mu_0 {\bf j = A = A_j + A}_a \,,
  \end{eqnarray}
where ${\bf A_j}$ is the vector potential of the supercurrent
density $j$ and ${\bf A}_a$ is the vector potential of the
applied field, e.g.\ ${\bf A}_a = -{\bf \hat z} x \mu_0 H$
for strips (${\bf \hat z}$ = unit vector along $z$).
Inverting $\mu_0{\bf j = - \nabla^2 A_j}$ we get for thick strips
  \begin{eqnarray}    
   A_j({\bf r}) = -\mu_0\! \int\! d^2r'\, {\ln |{\bf r-r}'|
        \over 2\pi} \, j({\bf r}')
  \end{eqnarray}
with ${\bf r}=(x,y)$, ${\bf A = \hat z} A$, ${\bf j = \hat z }j$.
The integration is over the strip cross section.
From Eqs.~(20,21) we have
  \begin{eqnarray}    
   A_a({\bf r}) = \mu_0\! \int\! d^2r'\,\Big[ {\ln |{\bf r-r}'|
   \over 2\pi} - \lambda^2 \delta({\bf r-r}') \Big]\,j({\bf r}') \,,
  \end{eqnarray}
Solving for $j$ and using $A_a = -x \mu_0 H$ we obtain
  \begin{eqnarray}    
   j({\bf r}) = H \int\! d^2r' K({\bf r,r}') \, x' \,, \\
   K({\bf r,r}') =  \Big[ -{\ln |{\bf r-r}'| \over 2\pi }
     + \lambda^2 \delta({\bf r-r}') \Big]^{-1} \,.
  \end{eqnarray}
The $\lambda$ dependent integral kernel $K({\bf r,r}')$ may be
computed by choosing ${\bf r}$ and ${\bf r}'$ on a grid and inverting
the resulting matrix similar as shown in Refs.\ \onlinecite{14,16}.

   From Eq.~(23) the current density induced by $H$ in the
Meissner-London state is obtained by a simple integration over
$x$ and $y$. Due to the symmetry of
$j(x,y)=j(x,-y)=-j(-x,y)=-j(-x,-y)$ it suffices to integrate
over a quarter of the strip cross section, $0 \le x \le a$,
$0 \le y \le b$, if the kernel is made symmetric.

Similar equations follow for the strip with current $I$,
  \begin{eqnarray}    
   j({\bf r}) =   I\! \int\! d^2r' K({\bf r,r}') \Big/ \!
      \int\! d^2r\! \int\! d^2r' K({\bf r,r}')
  \end{eqnarray}
with the same kernel, Eq.~(24), which however has a different
symmetric form since now $j(x,y)=j(x,-y)=j(-x,y)=j(-x,-y)$ holds.
For a strip with both applied field $H$ and current $I$ the two
solutions, Eqs.~(23,25) may be superimposed. The same method
works for thick disks in axial field if the appropriate kernel
$K$ is used, similarly as shown in Ref.~\onlinecite{16}.

   From the current density $j(x,y)$ the magnetic induction
${\bf B}(x,y) = (B_x, B_y) = (\partial A / \partial y,
 - \partial A / \partial x)$ is obtained via the Biot-Savart
law or by using Eq.~(21) to get $A(x,y)=A_j+A_a$.
In strip geometry the magnetic field lines are simply
the contour lines of $A(x,y)$. In the London case inside the
superconductor these field lines coincide with the contour
lines of the current density since $ A = -\mu_0 \lambda^2 j $,
see Figs.~3 and 4.

  Figures 3 and 4 show that the current density exhibits a
sharp finite peak in the corners; for the depicted case
$a=b=40\lambda$ this enhancement is $j(a,b)/j(a,0) = 3.7$
for $H>0$, and $j(a,b)/j(a,0) = 5.2$ for $I>0$. For small
$\lambda$, this enhancement increases as $\lambda^{-1/3}$.
Clearly, this
current peak favors the nucleation of vortex loops \cite{2}
at the corners and thus reduces the penetration field.
It may also enhance the nonlinear Meissner effect \cite{8,9}.

  Figure 5 shows the magnetic moment $M(a,b,\lambda)$ of
London strips with various aspect ratios $b/a$ as function
of $\lambda$ and normalized to the ideal Meissner moment
$M(a,b,0)$, Eq.~(6), and to the known $M(a,\infty,\lambda)$ of
the infinite slab. One has
 \begin{eqnarray}    
   \lim_{b \to \infty} {-M(a,b,\lambda) \over 4abH} =
    1 -{\lambda\over a}  \tanh{a \over \lambda} \,,
  \end{eqnarray}
with the limiting cases $1 - \lambda/a$ for $\lambda \ll a$
and $a^2 / (3\lambda^2) $ for $\lambda \gg a$. For thin strips
with $b \ll a$ and $b < \lambda$ we find the limits (see also
the Ohmic strip and disk in Ref.~\onlinecite{18})
  \begin{eqnarray}    
  {-M \over \pi a^2 H} = \left\{
  \begin{array}{r l}
    1 -{2 \lambda^2 \over \pi ab} \ln
    \left( 5.2{ab \over \lambda^2} \right), & ~\lambda^2 \ll ab\,,
                \\[.1 cm]
    {4ab \over 3\pi \lambda^2} - {2a^2 b^2
    \over \pi^2 \lambda^4 }+\dots\,, & ~\lambda^2 \gg ab\,.
  \end{array} \right.
  \end{eqnarray}
 Note the nonanalytic $\lambda$ dependence of $M$ for
small $\lambda$, which can be seen with the curve $b/a=0$
in Fig.~5.

   In summary, we found the exact analytical solution for the
magnetic moment and surface screening currents of long strips
with rectangular cross section in the ideal-screening Meissner
state generated by a homogeneous magnetic field $H$ and/or
transport current $I$. Accounting for a finite London penetration
depth $\lambda$, we present some explicit limiting expressions
and numerical results which show a high and sharp but finite
peak of the current density $j(x,y)$ along the four corners.
This sharp peak favors the penetration of magnetic vortices
from the corners, in form of quarter loops spanning
the corner. From the known $j(x,y)$ the exact shape and growth
of these loops can be obtained in principle, and thus both
the microscopic Bean-Livingston barrier \cite{2,3} and the
macroscopic geometric barrier \cite{4,5,6,7}, as well as the
thermally activated penetration of vortex loops from the corners
can be investigated in detail.

   Finally, this pronounced current peak is expected to enhance
the nonlinear Meissner effect predicted in $d$-wave
superconductors \cite{8} and to explain its dependence on the
sharpness of the specimen corners as observed in Ref.~\onlinecite{9}.

\vspace{-0.2 cm} 

 \begin{figure}[F1]
\epsfxsize= .95\hsize  \vskip 1.0\baselineskip
\centerline{ \epsffile{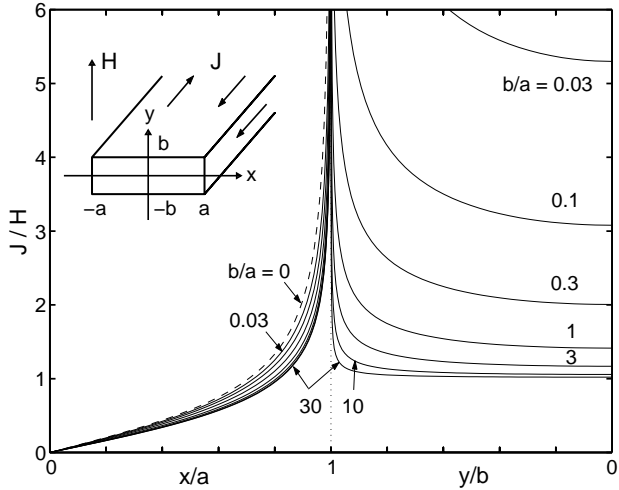}}
 \vspace{.1cm}
\caption{The Meissner surface currents $J(x,b)$ and $J(a,y)$,
Eqs.~(4,5), in strips with various aspect ratios $b/a$, see inset.
The dashed line gives the thin strip limit, Eq.~(12).
 } \end{figure}

 \begin{figure}[F2]
\epsfxsize= .95\hsize  \vskip 1.0\baselineskip
\centerline{ \epsffile{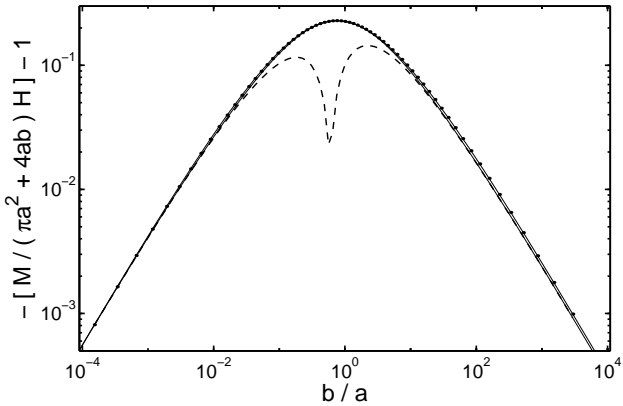}}
 \vspace{.2cm}
\caption{Magnetic moment of Meissner strips versus the
aspect ratio $b/a$, Eq.~(6) (solid line). The solid line with dots
shows the fit (8), and the dashed line the limits (9) and (10).
 } \end{figure}

 \begin{figure}[F3]
\epsfxsize= .99\hsize  \vskip 1.0\baselineskip
\centerline{ \epsffile{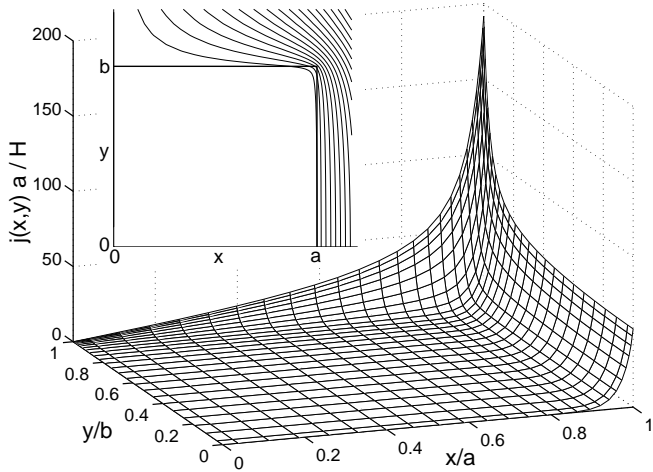}}
 \vspace{.1cm}
\caption{The current density $j(x,y)$ along a London strip with
square cross section ($a=b$) and London depth $\lambda/a=0.025$
in a perpendicular magnetic field $H$. A quarter of the
cross section is shown. Note the sharp but finite peak in the
corner. The inset shows the magnetic field lines.
 } \end{figure}

 \begin{figure}[F4]
\epsfxsize= .99\hsize  \vskip 1.0\baselineskip
\centerline{ \epsffile{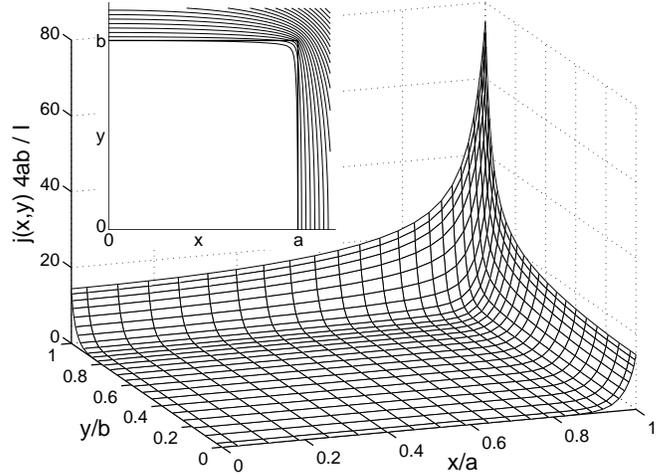}}
 \vspace{.1cm}
\caption{The current density $j(x,y)$ along the London strip
of Fig.~3 ($a=b=40\lambda$) but with transport current $I$
and no applied field, $H=0$.
 } \end{figure}

 \begin{figure}[F5]
\epsfxsize= .95\hsize  \vskip 1.0\baselineskip
\centerline{ \epsffile{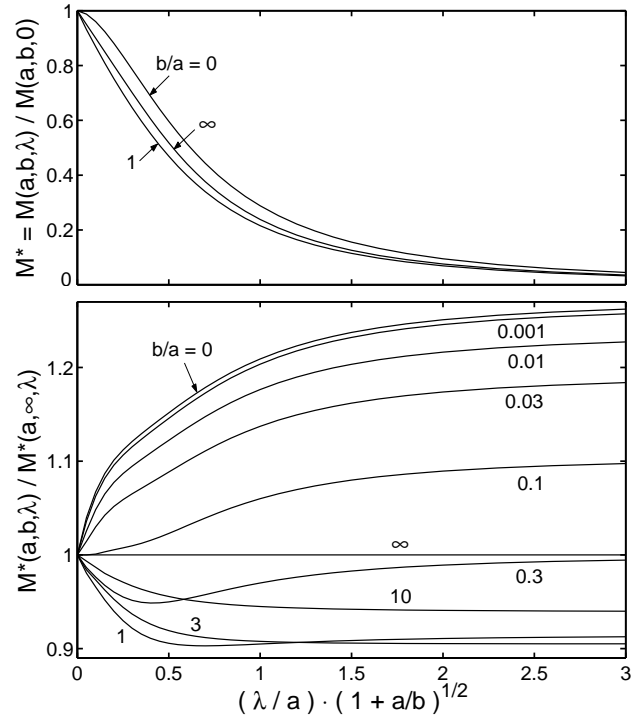}}
 \vspace{.1cm}
\caption{Reduced magnetic moment of London strips
$M^*=M(a,b,\lambda)/M(a,b,0)$ for various aspect ratios $b/a$.
Top:    Plotted versus a scaled London depth
        $(\lambda/a)\sqrt{1+a/b})$ the curves
        $M^*(a,b,\lambda)$ almost collapse.
Bottom: $M^*(a,b,\lambda)$ referred to the longitudinal limit
        $M^*(a,\infty,\lambda)$, Eq.~(26).
 } \end{figure}

\end{multicols}

\references
\vspace{-1.5 cm} 

\bibitem{1} F.~London, {\it Superfluids}, Vol.~I
            (Wiley, New York 1950).

\bibitem{2} C.~P.~Bean and J.~D.~Livingston, \prl{\bf 12},
            14 (1964);
            B.V.~Petukhov and V.R.~Chechetkin, Sov.\ Phys.-JETP
            {\bf 38}, 827 (1974);
            A.E.~Koshelev, Physica C {\bf 191}, 219 (1992);
            ibid. {\bf 223}, 276 (1994).
\bibitem{3} L.~Burlachkov, \prb{\bf 47}, 8056 (1993).

\bibitem{4} E.~Zeldov et al.,
           \prl{\bf 73}, 1428 (1994);
           Th.~Schuster et al.,   
           \prl{\bf 73}, 1424 (1994).

\bibitem{5}  N.~Morozov et al., \prl{\bf 76}, 138 (1996).

\bibitem{6} M.~Benkraouda and J.~R.~Clem, \prb{\bf 53}, 5716
            (1996);  \prb{\bf 58}, 15103 (1998).

\bibitem{7} E.~H.~Brandt, \prb{\bf 59}, 3369 (1999);
             ibid. {\bf 60}, 11939 (1999).

\bibitem{8} S.~K.~Yip and J.~A.~Sauls, \prl{\bf 69}, 2264 (1992).

\bibitem{9} C.~P.~Bidinosti et al.,
            \prl{\bf 83}, 3277 (1999).

\bibitem{10} L.~D.~Landau and E.~M.~Lifshitz, {\it Electrodynamics of
             Continuous Media} (Pergamon, London, 1959).

\bibitem{11} G.~M.~Mikitik and E.~H.~Brandt, \prb{\bf 60}, 592 (1999).

\bibitem{12} J.~R.~Clem et al.,
             J.\ Low Temp.\ Phys.\ {\bf 6}, 449 (1973).

\bibitem{13} K.~V.~Bhagwat and D.~Karmakar,
             Europhys.\ Lett.\ {\bf 49}, 715 (2000).    
\bibitem{14} E.\ H.\ Brandt, \prb{\bf 54}, 4246 (1996). 

\bibitem{15} E.~H.~Brandt, M.~V.~Indenbom, and A.~Forkl,
             Europhys.\ Lett.\ {\bf 22}, 735 (1993);
             E.~H.~Brandt and M.~V.~Indenbom, \prb{\bf 48}, 12093
            (1993); E.~Zeldov et al., \prb{\bf 49}, 9802 (1994).

\bibitem{16} E.~H.~Brandt, \prb{\bf 58}, 6506, 6523 (1998). 

\bibitem{17} R.~Prozorov et al.,
             \prb{\bf 83}, 115 (2000). 

\bibitem{18} E.~H.~Brandt, \prb{\bf 50}, 13833 (1994). 
\end{document}